\documentclass[%
 reprint,
superscriptaddress,
showpacs,preprintnumbers,
 amsmath,amssymb,
 aps,
 prc,
floatfix,
]{revtex4-1}

\usepackage[T1]{fontenc}
\usepackage[latin2]{inputenc}

\usepackage{bbold}
\usepackage{bbm}
\usepackage{color}
\usepackage{graphicx}
\usepackage{subfigure}

\usepackage{amsmath}
\usepackage{amssymb}
\usepackage{amsfonts}

\usepackage{siunitx}

\newcommand{\Ef}{ \mathcal{E} }
\newcommand{\Af}{ \mathcal{A} }

\newcommand{\Id}{ \mathrm{d} }
\newcommand{\IA}{ \mathrm{d^2} }
\newcommand{\IV}{ \mathrm{d^3} }

\begin{document}
\title{Anomalous baryon production in heavy ion collisions at LHC energies}
\author{Dániel Berényi}
\affiliation{Eötvös Loránd University, Budapest, Hungary}
\affiliation{MTA WIGNER RCP, RMI, P.O.B. 49, Budapest, 1525, Hungary}
\author{Attila Pásztor}
\affiliation{Eötvös Loránd University, Budapest, Hungary}
\author{Vladimir Skokov}
\affiliation{Brookhaven National Laboratory, Upton, USA}
\author{Péter Lévai}
\affiliation{MTA WIGNER RCP, RMI, P.O.B. 49, Budapest, 1525, Hungary}

\begin{abstract}
In ultrarelativistic heavy ion collisions the baryon-meson ratio is not fully understood
in the intermediate-to-high transverse momentum region. Although quark coalescence methods
combined with jet fragmentation yield results close to the experimental data, some tendencies
are not reproduced and require further investigation. We propose a new channel, namely extra quark-antiquark and diquark-antidiquark pair production from coherent gluon field created in the early stage of heavy ion collisions. This process is described quantitatively by the time-dependent Schwinger-mechanism. The extra diquark-quark coalescence yields are able to explain the measured anomalous baryon production. 

\end{abstract}
\maketitle

\section{Introduction}

The aim of ongoing investigations at ultrarelativistic heavy ion 
colliders e.g. the BNL Relativistic Heavy Ion Collider (RHIC), and the CERN Large Hadron Collider (LHC),
with up to $\sqrt{s}=\rm{200\ \rm{AGeV}}$ and $\sqrt{s}=\rm{5.5\ \rm{ATeV}}$ center of mass energies respectively, 
is to create an extremely high energy density, comparable to that of the early Universe.
At such high energies the colliding nuclei can be modeled by
two strongly Lorentz-contracted sheets of nucleons, with 
$\gamma \approx 100$ at RHIC and $\gamma \approx 2750$ at LHC, surrounded by a gluon cloud. Their
overlap creates a strong and rapidly changing color field. Particle (eg. quark-antiquark) pairs are produced 
in this field, a process similar to the Schwinger-mechanism in Quantum Electrodynamics (QED). 

The phenomenology of particle production in high energy $pp$ and heavy ion collisions traditionally
relies on the introduction of Quantum Chromodynamics (QCD) strings or ropes. The string picture provides a mechanism to convert
the collisional energy to field energy, which in turn is converted to particle pairs. Simple models
\cite{FRITIOF,LUND} are based on a time-independent picture and work properly at ISR and SPS energies. However
at extreme high energies, where the field amplitude is rapidly changing, like at LHC, this is not 
a realistic assumption. 

After creation, the particles start to interact with the surrounding matter, so
low-momenta light and strange quarks lose most of their original momentum, and thermalize. 
In contrast, high-momentum particles quickly escape from the strongly interacting matter, and maintain
their original momentum spectra. This way, the measured high momentum hadron spectra will be 
closely related to that of the primordial quarks. 

However, in a strong field, not only quark-antiquark pairs can be created, but composite
objects, such as diquarks too. We are focusing here on primordial diquark-antidiquark yields, 
as an extra source to final state hadron production at intermediate transverse momenta.

We build a simple phenomenological framework for the description of
mid-to-high transverse momenta hadron production in heavy ion collisions, focusing on
LHC energies. In addition to the usual thermal coalescence/recombination \cite{ALCOR1,ALCOR2,Scheibl:1998tk,hwa,greco,fries} and 
hard scattering (perturbative QCD) \cite{Brock:1994er, Owens:1986mp, Dokshitzer:1991wu,Yi02}, we consider new coalescence channels, 
that are based on a dynamical model of quark and diquark pair production in strong fields \cite{LevaiSkokovSpectra,StrongField10,QM2011}.

After reproducing the existing experimental data from the ALICE and CMS 
experiments on unidentified hadron yields,
we argue, that the inclusion of these new channels can lead to anomalous (enhanced) 
baryon-to-meson ratios in the mid-to-high transverse momentum range. 
Such anomalous ratios have already been observed at RHIC \cite{prpi_PHENIX1,prpi_PHENIX2,prpi_STAR,raa_prpi_STAR1,raa_prpi_STAR2}. From our model, 
we expect a similar effect at LHC, where the experimental data on identified
hadron spectra at such momenta is not yet available, but will be, after the accumulation
of sufficient statistics. 

This paper is structured in the following way. In the 2nd section we briefly review
the usual description of hadron production, including the thermal coalescence and the
jet fragmentation channels, and show our results obtained with them. The 3rd section 
reviews the traditional string phenomenology, followed by the Wigner-function based derivation 
of the time-dependent model of pair production. In the 4th section we present the numerical
results obtained by our framework. We close the paper with a brief discussion of our results,
 and an outlook to possible future investigations.

\section{Phenomenology of hadron production at mid-to-high $p_T$}
In proton-proton collisions, the hadron production is usually 
described by string fragmentation at \mbox{$p_T \lesssim 2 \ \rm{GeV/c}$}
and jet fragmentation at $p_T \gtrsim 2 \ \rm{GeV/c}$ \cite{Owens:1986mp}. In heavy ion collisions a 
large number of the produced partons are thermalized, creating 
the wanted quark-gluon plasma. These low momentum thermal quarks
can form hadrons with higher momentum. This process is usually
called quark coalescence \cite{ALCOR1,ALCOR2,Scheibl:1998tk,hwa,greco,fries}. This channel will dominate 
the medium transverse momentum spectra. The validity of this
picture was supported by RHIC data on proton-to-pion ratios \cite{prpi_PHENIX1,prpi_PHENIX2,prpi_STAR,raa_prpi_STAR1,raa_prpi_STAR2} and
the elliptic flow of identified hadrons \cite{eflow_PHENIX}. Another consequence 
of this dense parton matter is the appearance 
of jet energy loss \cite{GLV}.

In this section, we briefly review these aspects of hadron production 
in heavy ion collisions, especially at LHC energies. 

\subsection{Jet fragmentation}

In proton-proton collisions, the production of the most abundant hadrons can
be calculated by the QCD improved parton model \cite{Brock:1994er, Owens:1986mp, Dokshitzer:1991wu}.
At first, we determine the primary quark and gluon distributions. Later on, we fragment these 
high energy partons by means of fragmentation functions. In heavy ion collisions an additional
process, namely jet energy loss needs to be included due to the produced hot dense deconfined matter. 
 
In our calculation, we use a leading order pQCD framework, to make the transition to heavy ion 
collisions simpler. In A+B heavy ion collisions, we consider a geometrical 
superposition of nucleon-nucleon collisions, and include the effect of
jet energy loss by shifting the fragmentation function. 

The basic equations for calculating hadron spectra read:
\begin{widetext}

\begin{eqnarray}
\begin{array}{lcl}
E \frac{\IV \sigma}{\IV p} (pp \to h + X) = \sum_{abcd} \int \Id x_a \Id x_b \Id z_c  f_{a/p}(x_a,Q^2) f_{b/p}(x_b,Q^2) D_{h/c}(z_c,Q_F^2) 
\frac{\hat{s}}{\pi z_c^2} 
\left( \frac{\Id \sigma}{\Id \hat{t}} \right)^{ab \to cd}\delta(\hat{s}+\hat{t}+\hat{u})  \rm{,}
\end{array}
\end{eqnarray}

\begin{equation}
\begin{array}{lcl}
E \frac{\IV \sigma}{\IV p} (AB \to h + X) = \\
= \int \IA b \ \IA r T_A(r) T_B(|\vec{b}-\vec{r}|)  \sum_{abcd} \int \Id x_a \Id x_b
f_{a/A}(x_a,Q^2) f_{b/B}(x_b,Q^2) D_{h/c}(z_c^*,Q_F^2) \frac{1}{\pi z_c} \frac{z_c^*}{z_c} 
\left( \frac{\Id \sigma}{\Id \hat{t}} \right)^{ab \to cd} \rm{.}
\end{array}
\end{equation}

\end{widetext}

We use LO MSTW parton distribution function \cite{Martin:2009iq} and AKK fragmentation functions \cite{Albino:2005me}. In heavy
ion collisions $T_A(b) = \int \Id z \rho_A (b,z)$ is the nuclear thickness function normalized as usual $\int \IA b T_A(b) = A$. The
Woods-Saxon formula is applied for the nuclear density function $\rho_A(b,z)$. The integral in $b$ indicates the
nuclear overlap and the consideration of the Glauber geometry. Focusing on high momentum parton
production, we neglect nuclear shadowing, so the parton distribution
of the nuclei reads:
\begin{equation}
      f_{a/A}(x)=\left( \frac{Z}{A} f_{a/p}(x) + \frac{N}{A}f_{a/n}(x)\right) \rm{.}
\end{equation}

The average jet energy loss is taken into account with the simplified GLV formula \cite{GLV}:
\begin{equation}
    \Delta E= \frac{C_R \alpha_s}{N(E)} \left( \frac{L}{\lambda}\right)^2 \frac{\mu^2 \lambda}{\hbar c} \log\left(\frac{E}{\mu}\right) \rm{,}
\end{equation}
where $C_R$ is the Casimir-factor of the jet, $N(E)$ is a smooth function of energy, calculable with the
help of \cite{GLV}, $\lambda$ is the radiated gluon free path, 
$\mu^2 / \lambda \propto \alpha_s^2 \rho$ is a transport coefficient of the medium proportional to the parton density $\rho$,
and $L$ is the plasma thickness. We concentrate on midrapidity ($y=0$), where the jet transverse momentum is
shifted before fragmentation by the medium to $p_c^*=p_C-\Delta E$, which changes the $z_c$ parameter in the integrand to
$z_c^*=z_c/\left(1-\frac{\Delta E}{p_c}\right)$. Then the hard reaction scale is $Q=\xi p_T$ and the fragmentation scale is
$Q_F=Q/z_c^*$. The parameter $\xi$ will be fitted to match the $pp$ data, and the same value is used for the heavy ion calculations. 

By means of this model, we can describe the very high momentum hadron production from jets. Our numerical results will be described later. 

\begin{figure*}[th!]
\begin{minipage}{\columnwidth}
	\includegraphics[width=1.0\columnwidth,angle=0]{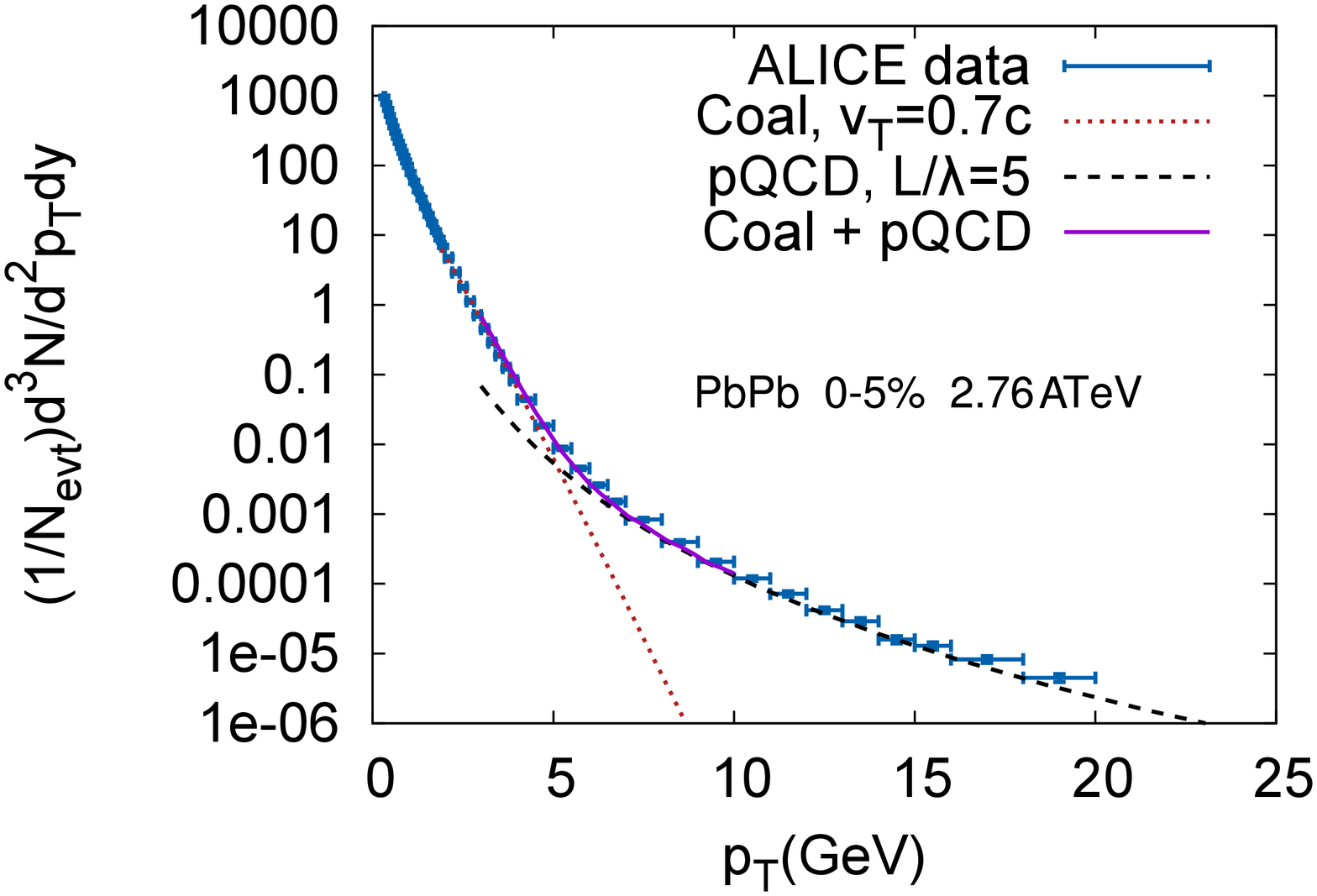}
	\caption{({\it Color online.}) Measured charged hadron transverse momentum spectra at $2.76\ \rm{ATeV}$
	in central $PbPb$ collisions \cite{ALICE_raa}. Calculation includes quark coalescence and 
	parton fragmentation yields with jet energy loss from GLV model~\cite{GLV} with
	opacity $L/\lambda=5$.}
\end{minipage}
\hfill
\begin{minipage}{\columnwidth}
	\includegraphics[width=0.68\columnwidth,angle=270]{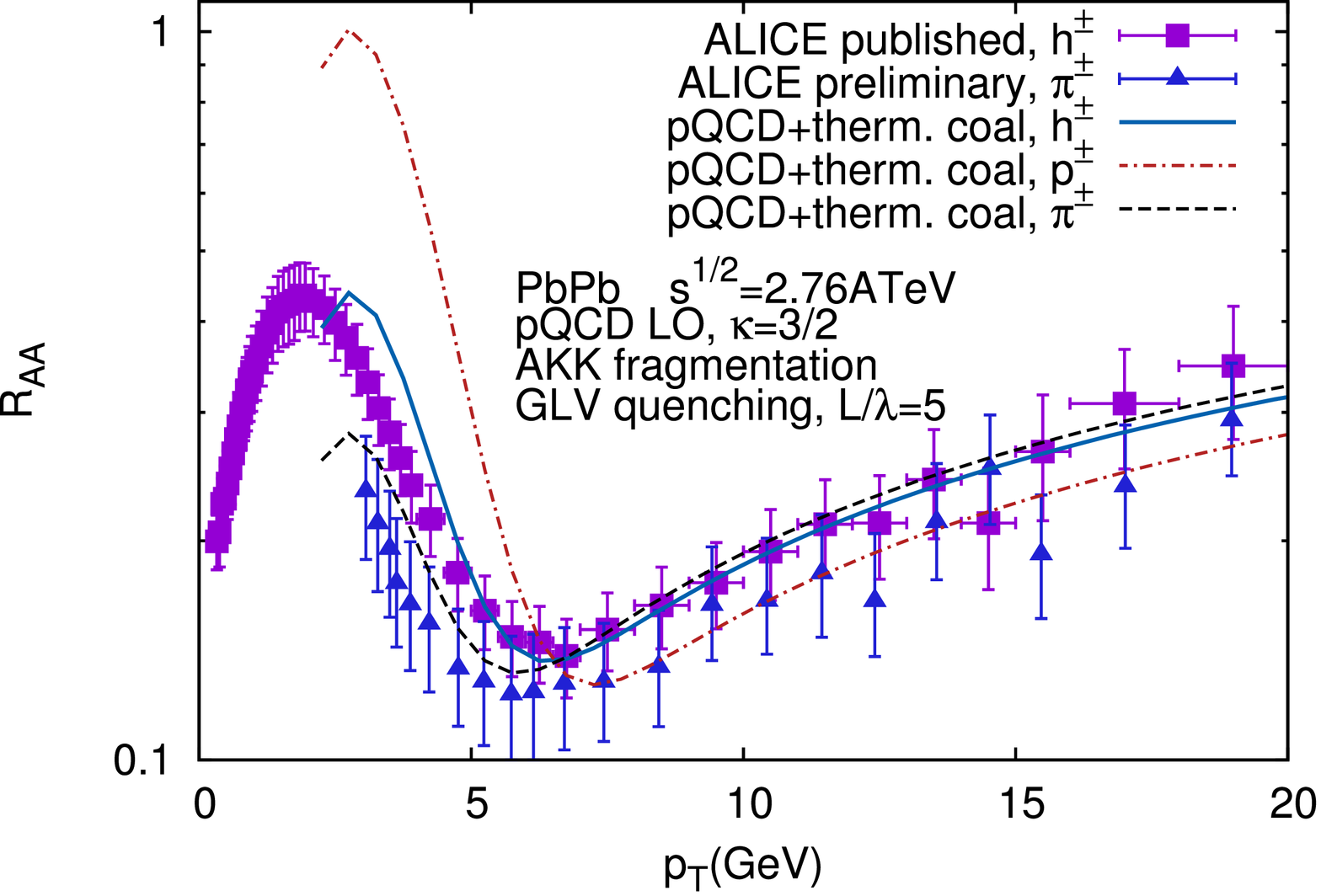}
	\caption{({\it Color online.}) Nuclear modification factor ($R_{AA}$) in a logarithmic scale
for charged hadrons, pions and protons according to our latest
calculations including quark coalescence and jet energy loss
 in central $PbPb$ collisions at $2.76\ \rm{ATeV}$. Data are from ALICE~\cite{ALICE_raa,ALICE_raa_qm11}.}
\end{minipage}
\end{figure*}

\begin{figure}[ht!]
	\includegraphics[width=0.68\columnwidth,angle=270]{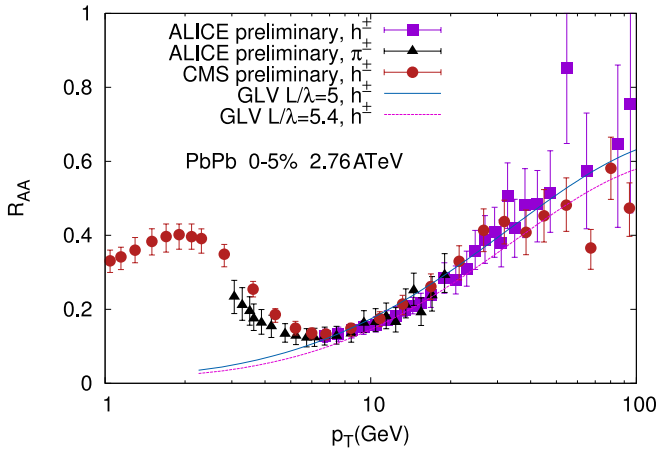}
	\caption{({\it Color online.}) Nuclear modification factor ($R_{AA}$) in a linear scale,
 obtained from simple perturbative-QCD 
calculations including jet energy loss in
central $PbPb$ collisions at $2.76\ \rm{ATeV}$, where the experimental data are from
ALICE~\cite{ALICE_raa_qm11} and CMS~\cite{CMS_raa_qm11}.}
\end{figure}

\begin{figure*}[t]
\begin{minipage}{\columnwidth}
	\includegraphics[width=0.74\columnwidth,angle=0]{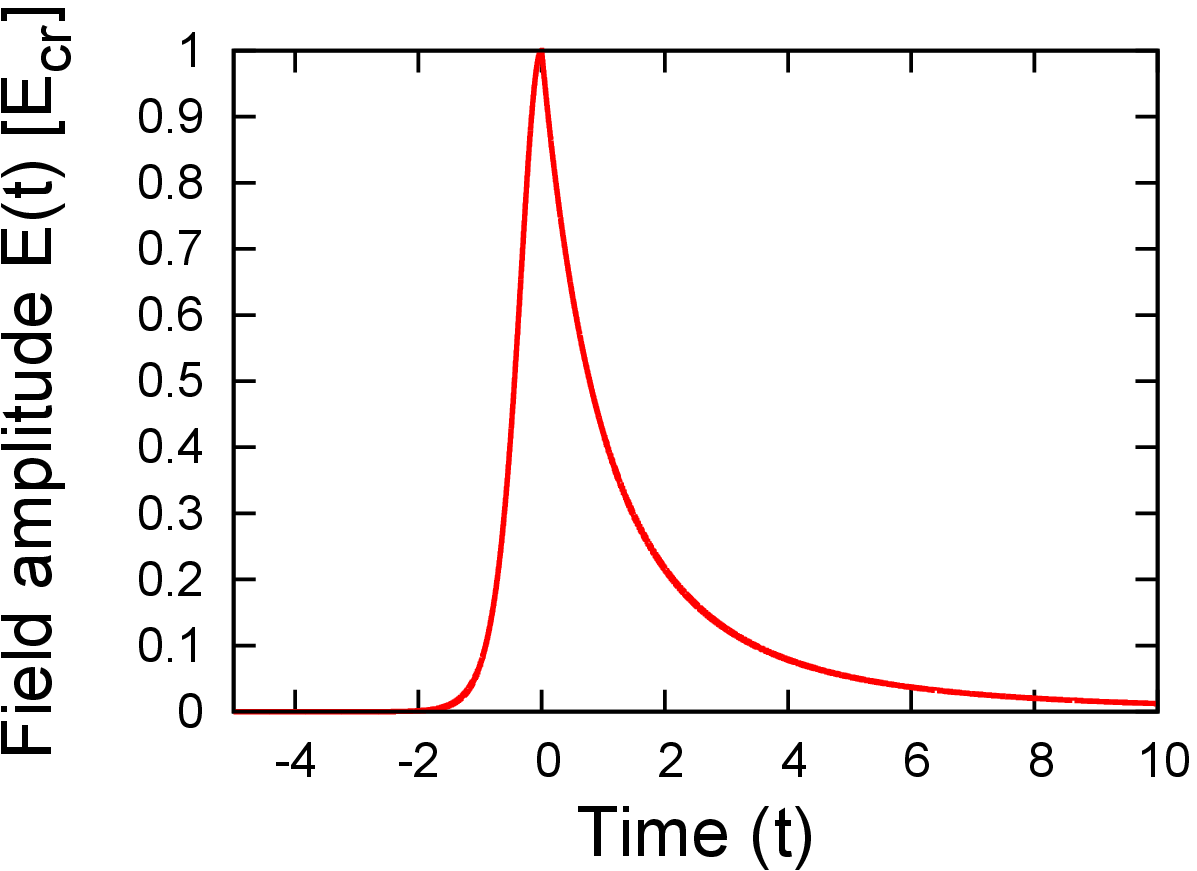}
	\caption{\protect\label{figField}({\it Color online.}) The external field used for the pair production channel, see eq. \ref{eq:external}.}
\end{minipage}
\hfill
\begin{minipage}{\columnwidth}
	\includegraphics[width=0.76\columnwidth,angle=0]{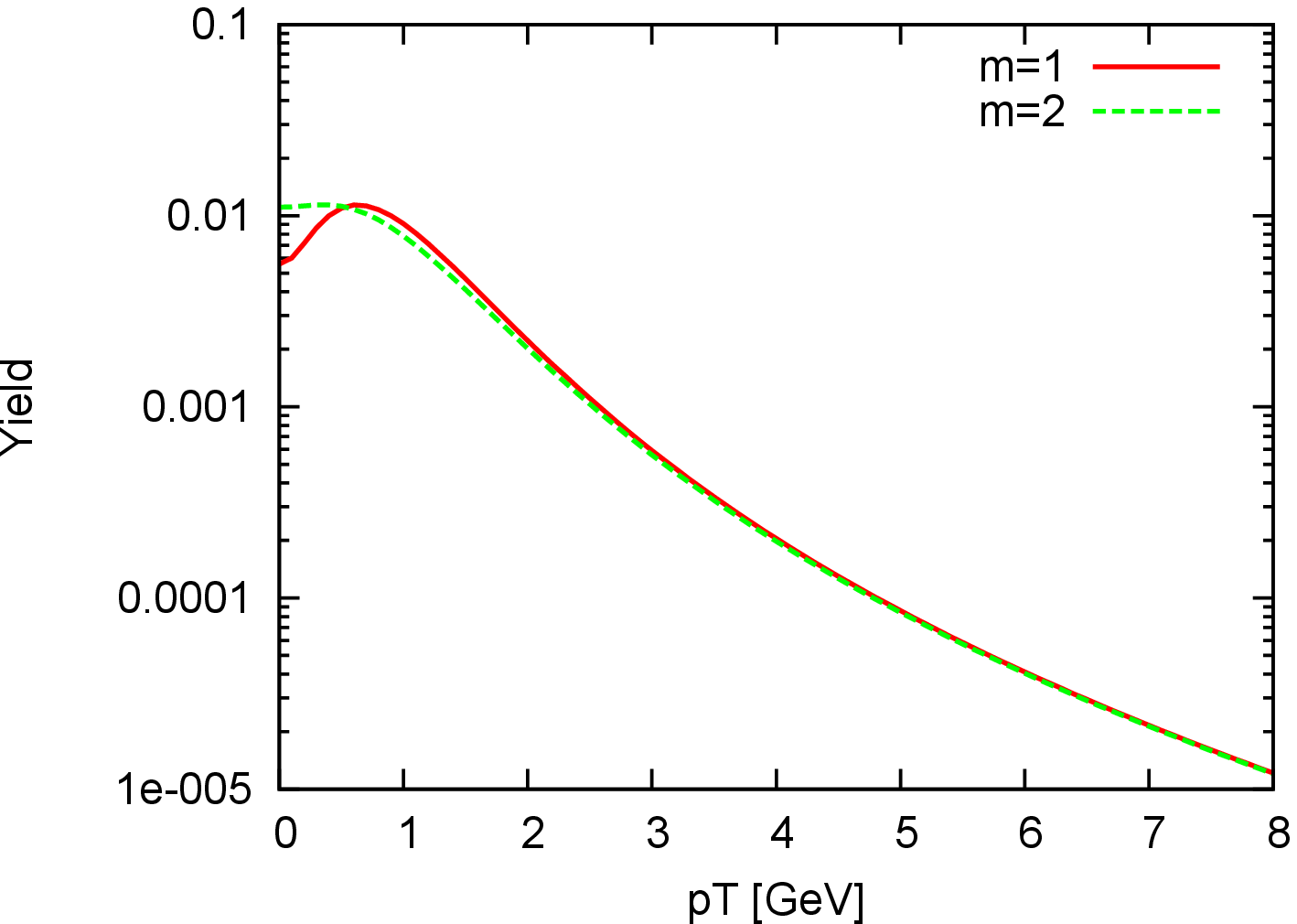}
	\caption{\protect\label{figSchwingerQ_QQ}({\it Color online.}) Particle spectra from the Schwinger-mechanism for quarks and diquarks.}
\end{minipage}
\end{figure*}

\subsection{Quark coalescence}

In the heavy ion collisions a dense parton matter will be produced. This matter
will thermalize and expand, reaching the deconfinement transition temperature known
to be $T \approx 180 \ \rm{MeV}$, calculated on the lattice \cite{lattice1,lattice2}.  At this point, 
the combination of participant quarks results in the dominant hadron production yield 
at low momenta, and give a significant
contribution in the intermediate momentum region, overlapping with jet
fragmentation. This yield will be determined by quark coalescence 
\cite{ALCOR1,ALCOR2,Scheibl:1998tk,hwa,greco,fries} applied in the low-to-intermediate transverse momentum 
range. Among the numerous models, we will use a simple, phase-space distribution based 
description. We assume an infinitely sharp (pre-)hadron momentum distribution, so the calculated
meson and baryon spectra reads:

\begin{eqnarray}
\label{eq:coal}
    E \frac{\IV N_{M}}{\IV P} & = & \frac{g_{M}}{(2\pi)^3} 
    \int_{\Sigma}P_{\mu} \Id \sigma^{\mu} f_a(r,\vec{P}/2) f_b(r,\vec{P}/2)  \rm{,} \nonumber\\
    &&\\
    E \frac{\IV N_{B}}{\IV P} &= & \frac{g_{P}}{(2\pi)^3} 
    \int_{\Sigma}P_{\mu} \Id \sigma^{\mu} f_a(r,\vec{P}/3) f_b(r,\vec{P}/3) f_c(r,\vec{P}/3) \rm{,} \nonumber\\
\end{eqnarray}

where $g_M$ and $g_P$ are the degeneracies of the resulting hadron, $g_{\pi^+}=1$, $g_{p^+}=2$, $g_{\rho^+}=3$, $g_{J^+}=3$ etc. The spacelike
hypersurface $\Sigma$ over which we integrate is chosen to be that of constant proper time $\tau=\sqrt{t^2-z^2}$, and the one
parton Wigner-functions are assumed to have the exponential form:

\begin{equation}
f_i(r,p)=\gamma_i e^{-pv/T} \theta(\rho_0-\rho) e^{-\frac{\eta^2}{2\Delta^2}} \rm{,}
\end{equation}

where the flow four-velocity $v$ follows a Björken scenario:
\begin{equation}
\begin{array}{rcr}
v = ( &\cosh\eta \ \cosh\eta_T, \\ &\sinh\eta_T \ \cos\phi, \\ &\sinh\eta_T \ \sin\phi, \\ &\sinh\eta \ \cosh\eta_T \ ) \rm{.}
\end{array}
\end{equation}
In this case the integral is completely insensitive to the parameter $\Delta$.
Essentially our free parameters will be the $\tau \rho_0^2 \pi = \tau A_T$ coalescence
volume, and the $v_T$ transverse flow velocity given by $\tanh \eta_T =  v_T$.
These two can be fitted to the $R_{AA}$ of unidentified hadrons, giving a prediction
for the identified spectra.

\subsection{The interplay of pQCD and coalescence}

Hadron production, especially charged pion, proton and antiproton
production have been measured with high precision at RHIC energies
in $pp$ and $AuAu$ collisions~\cite{prpi_PHENIX1,prpi_PHENIX2,prpi_STAR}. 
The analysis of the measured proton-to-pion ratio
proved the unsatisfactory performance of perturbative QCD
and independent jet fragmentation in the 
transverse momentum region of $2 \ \mathrm{GeV/c} < p_T < 8 \ \mathrm{GeV/c}$ ~\cite{Fai_PRL}. 
The introduction
of parton coalescence/recombination models~\cite{ALCOR1,ALCOR2,Scheibl:1998tk,hwa,greco,fries}
offered a possible solution and these calculations displayed the dominance 
of this channel in the intermediate-$p_T$ region. The resulting yields of our
model including fragmentation and coalescence is shown on Figure 1.
Figure 2. displays the nuclear modification factor, $R_{AA}$. Figure 3 shows
the extreme high momentum behavior of $R_{AA}$ obtained from a pure pQCD calculation.

However, the data on $R_{AA}$ at RHIC energies~\cite{raa_prpi_STAR1,raa_prpi_STAR2}  
displayed another anomaly: the $R_{AA}(p^+) > R_{AA}(\pi^+)$ in the window of $2 \ \mathrm{GeV/c} < p_T < 8 \ \mathrm{GeV/c}$
 and even beyond. 
This result contradicts the
expectations from jet energy loss descriptions (see e.g. the GLV-model~\cite{GLV})
and it was explained by quark-gluon jet conversion~\cite{qgjetconv1,qgjetconv2} 
and hadrochemistry inside jet-cones~\cite{hadrochem}.
Furthermore, a fine-tuned coalescence/recombination calculation may reproduce 
these data,
however, we would expect that this channel disappears at $p_T > 8 \ \mathrm{GeV/c}$ ~\cite{Levai08}. 
With the appearance of new ALICE data at LHC energies~\cite{ALICE_raa}, the question of
nuclear modification factors for identified hadrons appears again and it demands
a detailed analysis. In the following section we propose another process that may contribute to the aforementioned anomaly.

\section{Pair production in strong fields}
\subsection{Constant fields and string phenomenology}
Early models of hadron production in proton-proton collisions, e.g. the LUND and FRITIOF models \cite{FRITIOF,LUND}, 
were based on $q-\bar{q}$ and $qq-\bar{q}\bar{q}$ string formation. These strings are broken by the production
of new quark-antiquark pairs. These models suppose a time-independent color field, resulting in the $q-\bar{q}$ and $qq-\bar{q}\bar{q}$
yields:
\begin{equation}
\label{eq:string}
\frac{\rm{d}^7 N}{\rm{d} t \rm{d}^3 x \rm{d}^3 p} = 2 \exp\left(- \frac{\pi (m^2+\vec{p}_T^{\ 2})}{\kappa}\right)
\end{equation}

In this model, characterized by the $\kappa$ string constant, both the diquark and high momentum quark production
are suppressed.  

The description of the early stage of heavy ion collisions is different from the string formation in proton-proton collision, 
because of the higher density. For a very short time an extremely strong and rapidly changing color field will manifest. This
modification was mimicked by introducing color ropes \cite{BiroNielsen}, still described by equation (\ref{eq:string}), but
characterized by a larger string constant ($\kappa \gg 1\ \rm{GeV}/\rm{fm}$). However this will not change the basic suppression behavior. 
Recently, it has been shown, that introducing a flavour dependent string (rope) values the measured baryon-meson yields can be 
approximately reproduced \cite{ToporPop11}. We choose a different approach, relaxing the assumption of a constant field.

\begin{figure*}[th]
\begin{minipage}{\columnwidth}
	\includegraphics[width=0.68\columnwidth,angle=270]{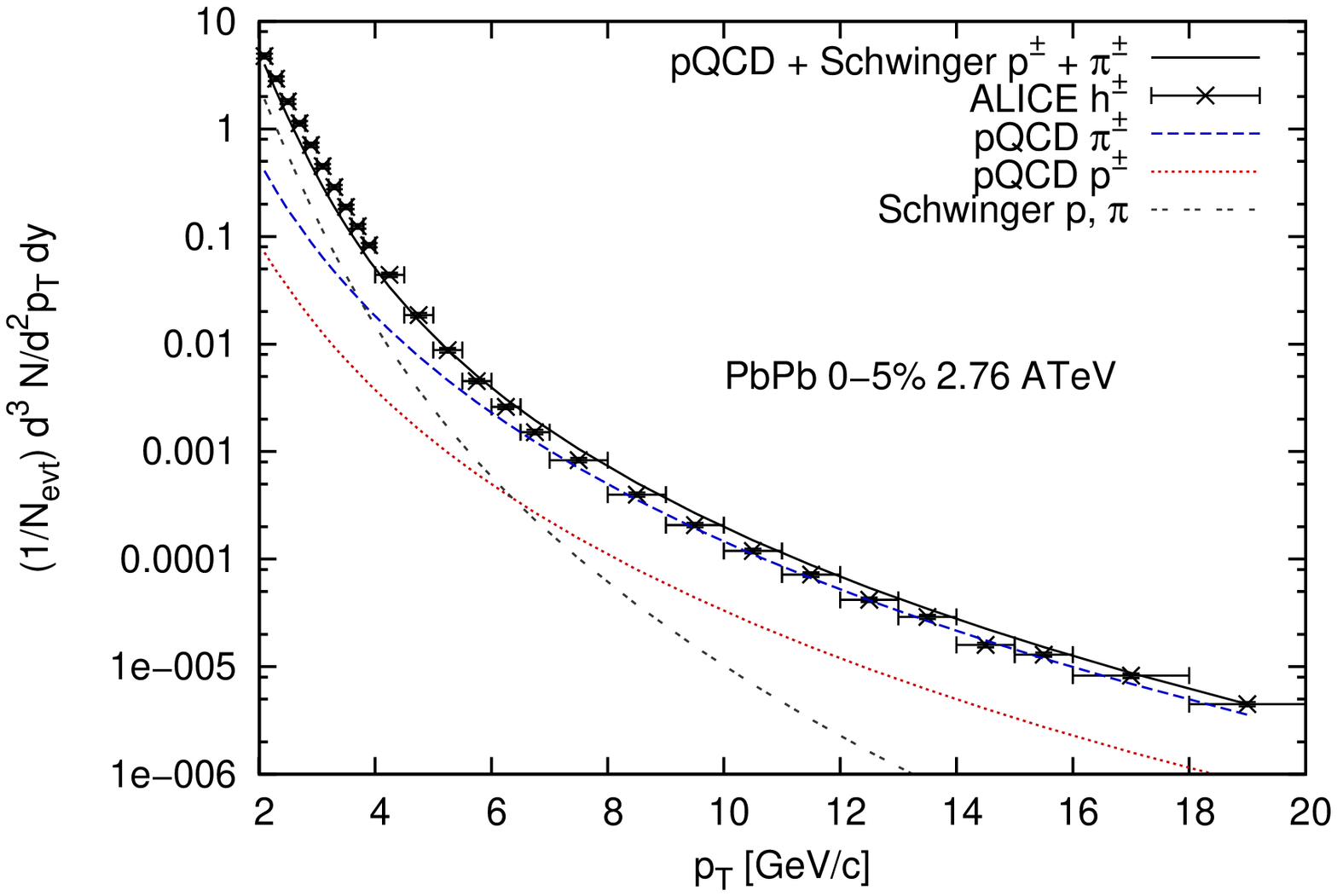}
	\caption{\protect\label{figFinalpQCD}({\it Color online.}) Anomalous proton and pion yields from strong color fields, comparing to
pQCD yields and ALICE data~\cite{ALICE_raa}.}
\end{minipage}
\hfill
\begin{minipage}{\columnwidth}
\includegraphics[width=0.68\columnwidth,angle=270]{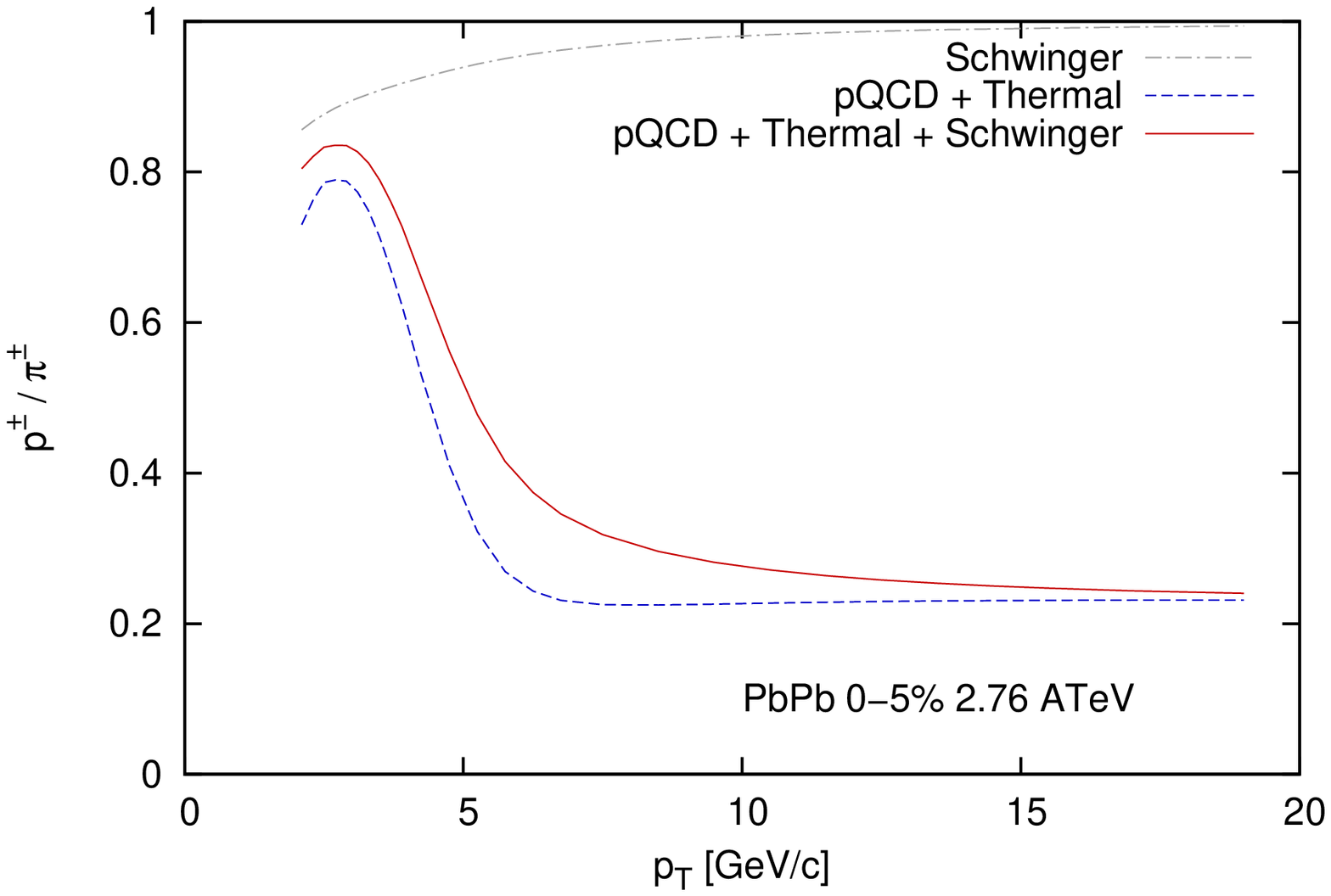}
	\caption{\protect\label{figp_per_pi}({\it Color online.}) The modification of the proton/pion ratio at LHC energies after introducing the
anomalous proton (and pion) yield connected to pair creation from strong field.}
\end{minipage}
\end{figure*}

\subsection{Time-dependent fields}
In this study, we use the Quantum Kinetic (QK) model for the description of pair production.
It was shown, that certain time-dependent external fields can result in power-law spectra \cite{LevaiSkokovSpectra}. The QK method can be extended to non-Abelian cases, but it was shown, that the behaviour has strong Abelian dominance \cite{LevaiSkokovSU2}. This is the basis of modeling the quark and diquark production from strong SU(3) fields with QED based calculations, because the spectra is expected to be the same except for the amplitude scale.

The quantum kinetic equation for the non-Abelian case can be derived from the time evolution of the corresponding Wigner function $W(\mathbf{k},t)$. The non-localities can be avoided by applying the gradient  approximation (see \cite{LevaiSkokovSU2} for details). The time evolution is then described by:
\begin{eqnarray}
&&\partial_t W+ \frac{g}{8}\frac{\partial}{\partial k_i}
\left( 4\{W,F_{0i}\} 
+ \right. \nonumber 
\\
&&\left. 
+2\left\{F_{i\nu},[W,\gamma^0 \gamma^\nu]\right\}-
\left[F_{i\nu},\{W,\gamma^0 \gamma^\nu\}\right] \right)=\nonumber\\
&&=ik_i \{\gamma^0 \gamma^i,W\}-im[\gamma^0,W] +ig \left[A_i\,, 
[\gamma^0 \gamma^i ,W]\right] . \ \ \ 
\label{KEW}
\end{eqnarray}
were $m$ denotes the mass of the particles in consideration, $g$ is the coupling constant, $A_\mu$ is the 4-potential of the external space-homogeneous color field and $F_{\mu\nu}$ is corresponding field tensor:
\begin{equation}
F_{\mu \nu} = \partial_\mu A_\nu - \partial_\nu A_\mu - i g  [ A_\mu, A_\nu ].
\end{equation}

While it is convenient to perform spinor and color decomposition of the Wigner function for direct numerical computations, if the symmetry of the external field is exploited, the kinetic equations can be recast in the form of the U(1) case, where the final equations for the particle density $f$ read~\cite{Prozorkevich:2004yp}: 
\begin{eqnarray}
\label{eq:qk}
	\frac{\Id f}{\Id t} & = & \frac{e\Ef \varepsilon_{\perp}}{\omega^2} v \\
	\frac{\Id v}{\Id t} & = & \frac{1}{2} \frac{e\Ef \varepsilon_{\perp}}{\omega^2} (1-2f) - 2 \omega u \\
	\frac{\Id u}{\Id t} & = & 2 \omega v.
\end{eqnarray}
Here $u$ and $v$ are auxiliary functions and
\begin{equation}
	 \omega^2(\vec{q}, t) = \varepsilon^2_{\perp} + \vec{q}^{\ 2}_{\parallel}
\end{equation}
\begin{equation}
	 \varepsilon_{\perp}^2 = m^2 + \vec{q}^{\ 2}_{\perp}
\end{equation}
\begin{equation}
	 \vec{q} = (\vec{p}_{\perp}, p_{\parallel} - e\Af(t))
\end{equation}

The initial conditions at $t=-\infty$ are $f(\vec{q})=v(\vec{q})=u(\vec{q})=0$. The only input for the QK model is 
the electric field $\Ef(t)$ and its vectorpotential $\Af(t)$. We chose the following time dependence
(see Fig. \ref{figField}):

\begin{equation}
\label{eq:external}
	 \Ef(t) = \begin{cases} E_0 \left( 1 - {\rm tanh}^2\frac{t}{\tau_1} \right), & t \leq 0 \\ E_0 \left( 1 + \frac{t}{\tau_2}\right)^{-\delta}, & t > 0. \end{cases}
\end{equation}
where we use the phenomenological field parameters ($E_0$, $\delta$, $\tau_1$, $\tau_2$) according to our expectations: the first part of the formula is motivated by the increasingly overlapping nucleons, avoiding a full self-consistent
QCD description. The decay of the field is suggested by Björken hydrodynamics.  The time-scale of the raise and decay
is $\tau_1$ and $\tau_2$ respectively. The field amplitude is measured in critical Schwinger field units: $E_{cr}=m^2_q / g$,  
set by the effective quark mass. We chose $m_q=300$ GeV. The QK equations are integrated numerically in time and in the longitudinal direction, 
giving $f(p_T)$ at $t=+\infty$ for quark and diquarks. Figure \ref{figSchwingerQ_QQ} illustrates the 
obtained $q$ and $qq$ spectra. Note, 
that at high momenta, the diquark and quark spectra are similar, and both are enhanced compared to the
case of a constant field. 

The input of the coalescence equations (\ref{eq:coal}) will then be:
\begin{equation}
	 f_S(r,p) = f(p_T)\Theta(\rho_0 - \rho)\exp\left(-\frac{\eta^2}{2\Delta^2}\right)\rm{,}
\end{equation}
where the parameters $\rho_0$ and $\Delta$ are the same as in the calculation of the thermal spectra, with the addition of a diquark + quark $\to$ hadron channel, governed by equations
formally equivalent to quark + quark $\to$ meson coalescence, only with a quark distribution interchanged with the diquark one.

\begin{figure*}[th]
\begin{minipage}{\columnwidth}
\includegraphics[width=0.68\columnwidth,angle=270]{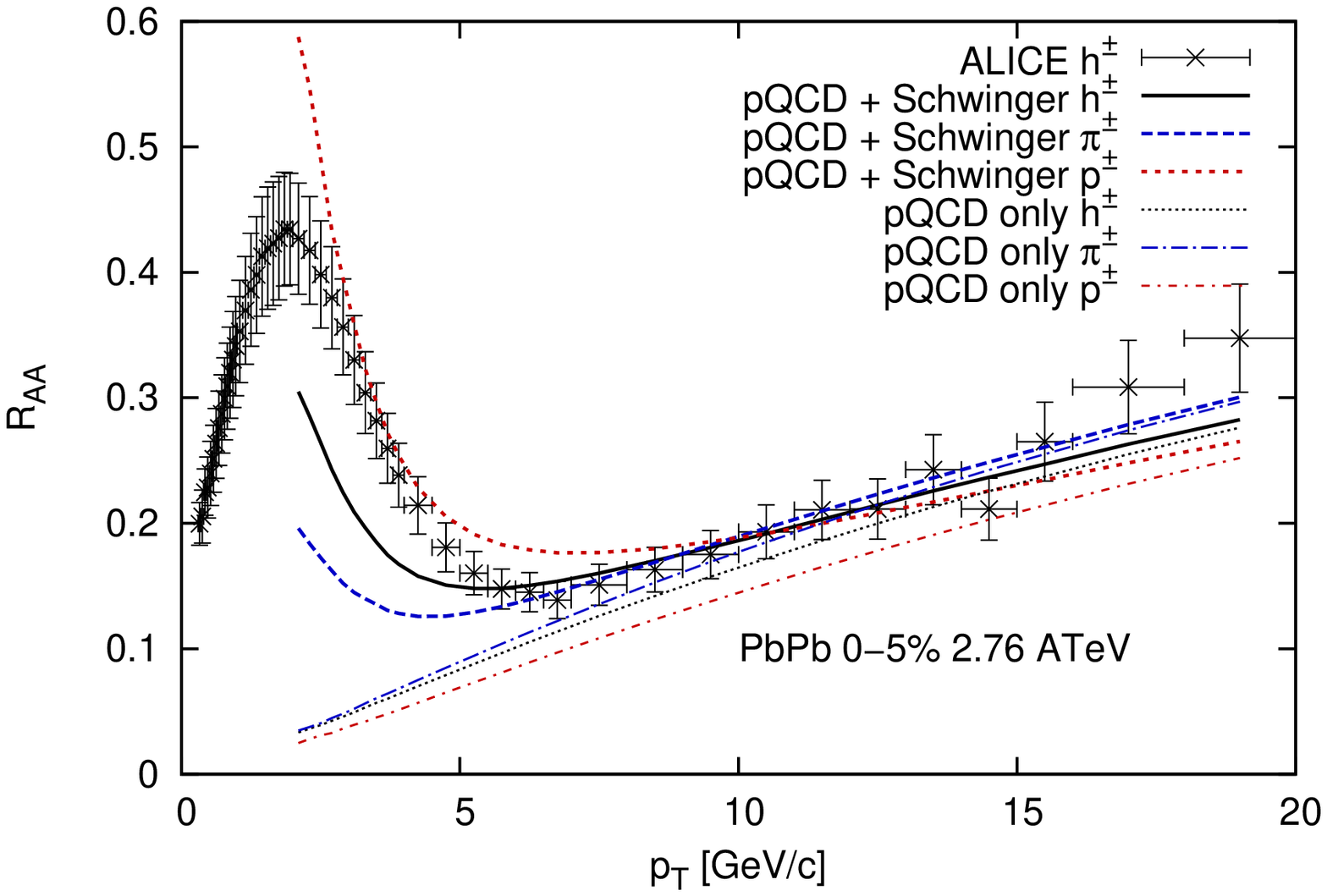}
	\caption{\protect\label{figFinalRAA}({\it Color online.}) The modification
of the nuclear modification factor ($R_{AA}$) connected to the
appearance of the anomalous hadron yields in PbPb collisions at
2.76 ATeV, compared to ALICE data~\cite{ALICE_raa}.}
	
\end{minipage}
\hfill
\begin{minipage}{\columnwidth}
	\includegraphics[width=1.0\columnwidth,angle=0]{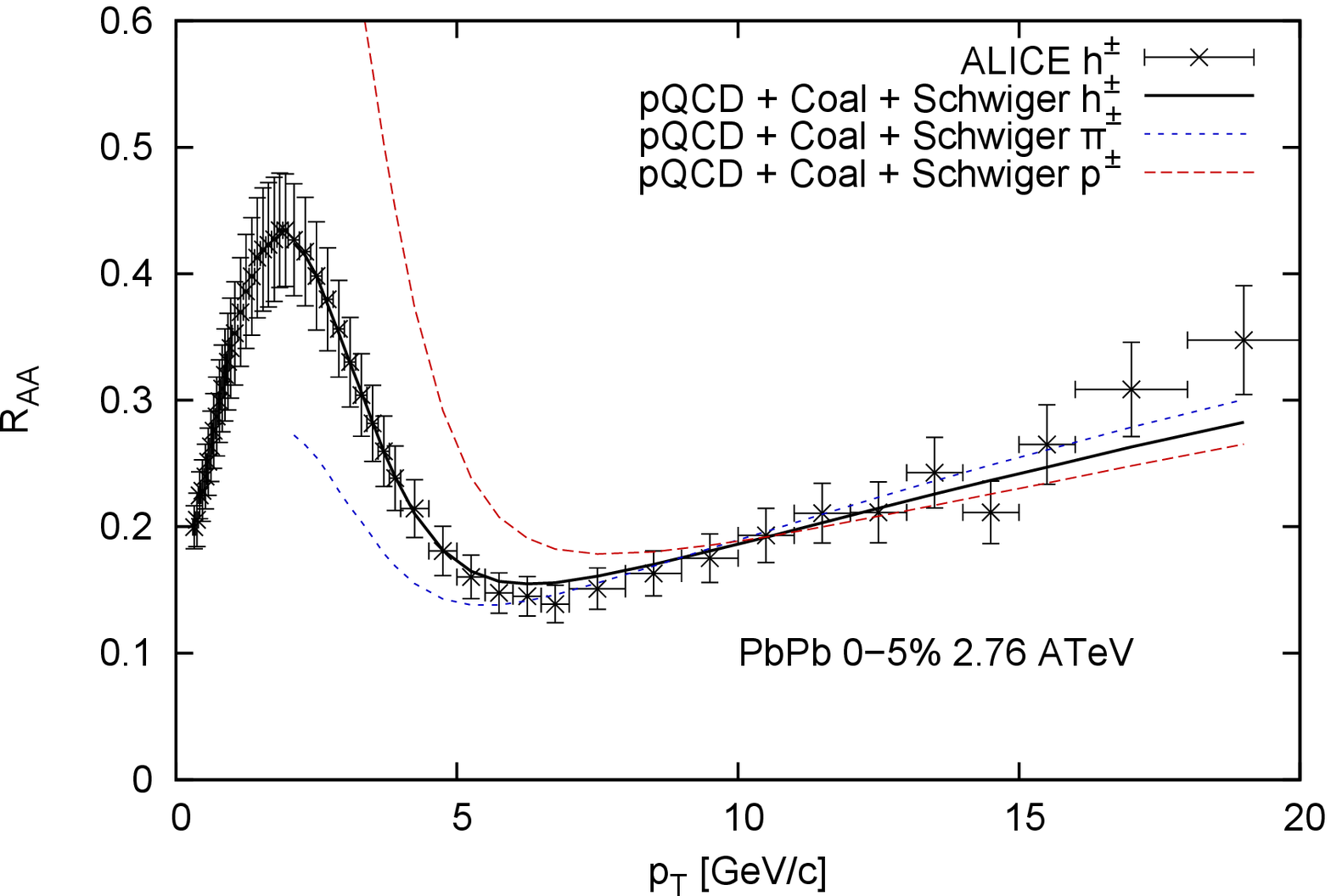}
	\caption{\protect\label{figFinal}({\it Color online.}) The nuclear modification factor ($R_{AA}$) after including 
jet fragmentation, quark coalescence and the yield from time-dependent strong coherent field. The theoretical results
are compared to  ALICE data~\cite{ALICE_raa}.}
\end{minipage}
\end{figure*}

\section{Numerical Results}
Earlier studies of particle production from time-dependent strong coherent 
fields~\cite{LevaiSkokovSU2,StrongField10} indicated that in case of a rapidly changing field, the obtained spectra of quark-antiquark pairs
could easily become power law, instead of the Gaussian drop of
usual Schwinger spectra from static strong electric field~\cite{ToporPop11}.
Thus we have an extra channel of quarks and antiquarks with momentum spectra
different from usual pQCD results. Furthermore, diquark-antidiquark pairs
can be created, opening an anomalous channel for baryon production. 
Since quark and diquark yield are very close to
each other in a time dependent field, the proton/pion ratio would be close
to unity for hadron coalescence from pair-creation channels.
Figure \ref{figSchwingerQ_QQ} displays these effects.
This anomalous yield of protons  from diquarks 
can overwhelm the fragmentation component, as can be seen on Figs. \ref{figFinalpQCD} and \ref{figFinalRAA}, 
where only the pQCD and pair creation channels are included.
The parameters of the time dependence ($\tau_1, \tau_2, \delta$) and the strength
of the electric field ($E_0$), will determine the resulting quark and diquark spectra. In our fit on the unidetified hadron $R_{AA}$ \cite{ALICE_raa} we found $E_0 = 2.2 E_{cr}$, $\delta = 6.0$, $\tau_1 = 0.31/m_q$, $\tau_2 = 0.36/m_q$ for the external field, and $\tau A_T = 4226\ \rm{fm}^3$ and $v_T = 0.7$ for the thermal coalescence parameters.

Finally, Fig. \ref{figFinal} displays  $R_{AA}$ obtained from the
combined framework, including parton fragmentation and quark coalescence using thermal source 
and pair production in time-dependent strong fields.  In this way we can approximately reproduce the measured nuclear modification factor
in central $PbPb$ collisions at $2.76\ \rm{ATeV}
$, and
a modified proton and pion suppression pattern will appear,
namely $\rm{R_{AA}}(\rm{proton}) > R_{AA}(\rm{pion})$, as shown on Fig. \ref{figFinal}). 
Naturally, this anomalous proton yield
will modify  the proton-to-pion ratio, although the modification could remain
relatively small, as indicated by Figure \ref{figp_per_pi}).

\section{Conclusions}
We have studied hadron production in heavy-ion collisions focusing on baryon-to-meson ratios and proton, pion $\rm{R_{AA}}$. We used a model combining jet fragmentation with jet quenching and coalescence in a thermal
quark matter characterized by the deconfinement transition temperature $T\approx180 \ \rm{MeV}$. We also introduced a
new channel into the coalescence model, originating from extra quark and diquark yields from
strong time-dependent fields. We have demonstrated that such a model can reproduce existing 
experimental data at LHC energies, and that an anomalous production, formerly seen at 
RHIC \cite{raa_prpi_STAR1,raa_prpi_STAR2}, could also be seen in future LHC data. 

High precision measurements 
of identified charged hadrons at LHC energies could become the basis
of a proper discussion on the formation of coherent field and the
existence of anomalous quark and diquark production channels. 
 
Future numerical studies could include a more realistic time and space dependence of the external gluon field
(see e.g.~\cite{Kovner:1995ja,Krasnitz:1998ns}) which may modify the spectra of produced quarks and diquarks. A similar investigation was discussed
for the Abelian case in Ref. \cite{BerenyiLEI}. \\

\section*{Acknowledgments}
This work was supported in part by Hungarian OTKA Grants No. 77816 and No. 106119.
The research of V.S. was supported by the US Department of Energy
under Contracts DE-AC02-98CH10886.

\end{document}